# Quantum Computing and Cybersecurity in Accounting and Finance: Current and Future Challenges and the Opportunities for Securing Accounting and Finance Systems in the Post-Quantum World

**Huma Habib Shadan**

humashadan@hotmail.com

Victoria University

and

**Sardar M.N. Islam**

Victoria University

## Abstract:

Quantum computing is transforming the world profoundly, affecting businesses, organisations, technologies, and human beings' information systems, and will have a profound impact on accounting and finance, particularly in the realm of cybersecurity. It presents both opportunities and risks in ensuring confidentiality and protecting financial data. The purpose of this article is to show the application of quantum technologies in accounting cybersecurity, utilising quantum algorithms and QKD to overcome the limitations of classical computing.

The literature review reveals the vulnerabilities of the current accounting cybersecurity to quantum attacks and the need for quantum-resistant cryptographic mechanisms. It elaborates on the risks associated with conventional encryption in the context of quantum capabilities. This study contributes to the understanding of how quantum computing can transform accounting cybersecurity by enhancing quantum-resistant algorithms and using QKD in accounting.

The study employs PSALSAR systematic review methodology to ensure rigour and depth. The analysis shows that quantum computing enhances encryption techniques to superior possibilities than classical. Using quantum technologies in accounting minimises data breaches and unauthorised access. The study concludes that quantum-resistant algorithms and quantum key distribution (QKD) are necessary for securing the accounting and finance systems of the future.

# 1.0 An Introduction: Quantum Computing and Accounting and Finance System Cybersecurity Challenges:

## 1.1 Background of the Study:

Quantum computing and cybersecurity are colliding to redefine the future of digital trust in accounting and finance systems. The advent of quantum computing presents both an unprecedented threat and an extraordinary opportunity for the cybersecurity landscape of accounting and finance. Cybersecurity has become a crucial component of modern accounting and finance systems, driven by the growing sophistication of cyber threats. Traditional cryptographic techniques, such as RSA and Elliptic Curve Cryptography (ECC), face new vulnerabilities as financial data and transactions become more digitalised, particularly with the advancements in quantum computing.[1]. Financial organisations are particularly vulnerable to attacks from adversaries utilising quantum technology, as they primarily depend on public-key cryptography protocols. Quantum computing has the potential to transform current cybersecurity systems, making the accounting and finance industries particularly vulnerable to significant changes. By examining the relationship between quantum computing and the protection of accounting and financial systems, it becomes evident that this new technology is both a powerful ally and a formidable challenge. The article aims to elucidate the complexities of how quantum computing can enhance the cybersecurity of financial and accounting systems. The risks of transitioning from classical to quantum paradigms are as substantial as the promise of improved data protection and transaction security.

By conducting a thorough literature review, we aim to enhance our understanding of how quantum computing could transform cybersecurity in accounting processes. The findings indicate significant advancements in encryption methods, with quantum computing providing enhanced defences against online attacks. However, the journey toward integrating quantum technology is not without challenges, as it requires substantial modifications to current cybersecurity strategies. The investigation underscores the pressing need to develop quantum-resistant algorithms and implement Quantum Key Distribution (QKD) to secure financial data in the long term. As we navigate this technological transition, our goal remains to create more reliable and secure accounting systems, ensuring the accuracy and privacy of financial information in an increasingly digital world.

## 1.2 Motivation and Relevance:

The convergence of quantum computing (QC) and cybersecurity (CS) marks a pivotal turning point in the evolution of accounting and financial systems. With the increasing digitisation of financial operations, cybersecurity has become fundamental in protecting sensitive accounting data, client records, and strategic financial information [2]. However, existing cryptographic frameworks, such as RSA and Elliptic Curve Cryptography (ECC), which form the backbone of current cybersecurity protocols, are highly susceptible to quantum algorithms—namely, Shor's algorithm for integer factorisation and Grover's algorithm for search—which could render these methods obsolete [3]. Concurrently, quantum computing offers transformative computational advantages, enabling the high-speed resolution of complex financial problems, such as portfolio optimisation, risk modelling, and credit scoring [4]. However, the same computational power introduces new cybersecurity challenges, threatening the integrity of encrypted systems [5]. This duality of opportunity and risk positions the topic as both timely and urgent, offering the potential to conceptualise and implement quantum-resilient cybersecurity architectures within accounting frameworks, thereby paving the way for future-proof financial infrastructures.

### 1.3 Research Aim, Research Questions and Research Objectives:

#### 1.3.1 Research Aim:

The primary aim of this article is to explore and develop a comprehensive understanding of how quantum computing technologies can be applied to enhance the cybersecurity of accounting systems. The article aims to scrutinise the potential of quantum computing to strengthen cybersecurity measures in accounting, focusing on how quantum algorithms and technologies can address the limitations of classical computing in safeguarding sensitive financial data and ensuring secure transactions. It would not be wrong to say that this involves analysing the potential of quantum-resistant algorithms and Quantum Key Distribution (QKD) in safeguarding financial data against emerging cyber threats, thereby ensuring the integrity, confidentiality, and reliability of accounting processes in the quantum era.

#### 1.3.2 Research Question:

**a. Primary Question:**

The primary question that needs to be considered in this article is:

"How can quantum computing enhance CS in accounting and finance within an organization?"

It also compares the effectiveness of quantum-resistant algorithms in safeguarding accounting data through quantum-enabled attacks with classical cryptographic methods. This research also explores the response to how the conceptual model, based on the Thompson model [6], can be adapted to incorporate QC Technologies for enhanced customer satisfaction (CS) in accounting. Further research identifies the main challenges an organisation faces in adopting QC Technologies for improved cybersecurity in accounting and suggests ways to resolve them.

**b. Secondary Research Questions:**

**RQ1:** What are the critical computing threats in current accounting systems that quantum computing can address?

**RQ2:** What are the most promising quantum algorithms for securing financial data in accounting?

**RQ3:** How can quantum computation be integrated into current accounting systems without disrupting operations?

These research aims, objectives, and questions are designed to guide the exploration and analysis of quantum computing's role in transforming the cybersecurity landscape for accounting systems, ultimately contributing to the development of more secure and resilient financial infrastructures.

#### 1.3.3 Research Objectives:

The article clarifies the role of quantum computing as both a threat and a tool in accounting cybersecurity, responding to the aforementioned research questions. The objectives aligned with these questions include:

**a. Objective 1:**

Identify and analyze the major cybersecurity threats facing contemporary accounting and finance systems. Determine which threats could be mitigated or neutralised by quantum computing capabilities, which will be addressed through responding to Research Question 1.

b. **Objective 2:**

Investigate and evaluate quantum computing algorithms and techniques (such as quantum cryptography and quantum-resistant algorithms) that show promise for protecting financial data. Assess their applicability and effectiveness in accounting contexts, which is addressed by responding to Research Question 2.

c. **Objective 3:**

Develop a framework or guidelines for integrating quantum computing-based security measures into existing accounting information systems. The goal is to implement these measures to minimise disruption to business operations, which is addressed by responding to Research Question 3. It includes considering the organisational, technical, and regulatory factors that influence successful implementation.

### 1.4   Significance of Study:

This research is significant for both academia and industry. It is a unique study in which thorough research has not been conducted before. Adopting this research article academically, it bridges a gap between the domains of quantum computing and accounting information systems security, two fields that are rarely considered. It contributes to the theoretical understanding of how emerging technologies can transform cybersecurity practices in specific contexts (here, financial and accounting systems). Practically, the findings will help organisations in the financial sector prepare for the quantum disruptive effects. For instance, a recent industry survey found that over 70% of large U.S. firms expect quantum computers to break current encryption by 2030 and are "extremely concerned" about this threat [6]. However, many firms have not started planning adequate defences [6]. By providing insight into quantum-resilient security measures and integration strategies, this article offers timely guidance to practitioners (CISOs, IT managers, auditors, and accountants) on strengthening their systems before quantum attacks become a reality.

This study makes a significant transdisciplinary contribution to cybersecurity, financial systems engineering, and quantum information science. Establishing a robust framework for evaluating and implementing quantum cybersecurity techniques adds to the existing body of knowledge. It provides financial organisations with valuable insights for preparing their data systems against post-quantum threats. Additionally, the research addresses policy-level implications, emphasising the urgent need for industry leaders and regulatory agencies to prioritise investments in quantum-secure infrastructures and the upskilling of the cybersecurity workforce.

## 2.0   Literature Review and Theoretical Foundations:

### 2.1   Overview of Cybersecurity in Accounting and Finance:

The digital transformation [4] in the accounting and finance sectors has transformed traditional practices, offering enhanced operational efficiency, improved data accessibility, and advanced analytical capabilities. Digitalizing accounting records, electronic financial transactions, and cloud-based storage systems has significantly expanded the attack surface for cybercriminals. The researcher [7] emphasises the importance of comprehending cybersecurity risks fully when implementing digital transformation (DT) to prevent disruptions caused by malicious activity or unauthorised access by attackers seeking to modify, destroy, or extract sensitive data from users. Cybersecurity is essential to DT, providing security against online threats. To mitigate cybersecurity risks during digital transformation (DT) implementation, the research aims to investigate the impact of cybersecurity and digital transformation on business resilience. Prioritising cybersecurity measures and ensuring systems are safe from potential threats are essential for organisations that utilise digital transformation (DT) [8, 9]. The author adopted the PRISMA methodology and the systematic literature review approach, focusing on various sectors, including finance, healthcare, government, and industry, to discuss the unique challenges and solutions each faces in cybersecurity. Because business organisations are so diverse, the researcher explained in his article that their technology systems are also diverse. Therefore, organisations can enhance cybersecurity using contemporary technologies, such as the Internet of Everything [10]. [11] They presented theoretical and doable suggestions for improving economic security in the digital economy. When organisational procedures are converted into IT solutions, data transformation (DT) occurs. It can have a significant impact on various areas of a company.

However, [7] emerging technologies like blockchain, cloud computing, big data and analytics, artificial intelligence, and artificial intelligence pose cybersecurity risks as firms worldwide undertake digital transformation. New technologies pose unforeseen threats, such as cyberattacks, even as they help firms become more efficient and competitive [9]. Appropriate security measures should also be in place to protect technology infrastructures from cyberattacks [12]. Organisations can defend themselves against recurring cyberattacks by implementing a basic security plan [13]. The review necessitates proactive cybersecurity measures and research from a global perspective across various regions. It also suggests that organisations analyse cybersecurity investments to determine where frameworks could be developed.

### 2.1.1 Cybersecurity:

[14] The researcher in this article discusses cybersecurity, clarifying that it involves maintaining integrity, availability, and confidentiality. It also requires full authentication and non-repudiation to prevent harm to electronic communication systems and services. CISCO has explained cybersecurity and deployed several levels of security across networks and systems to thwart attacks on confidential data or commercial processes.

[15] The researcher explains in his article how Quantum computers are going to have an impact on cybersecurity concerns. The researcher discusses the study of cybersecurity in their article, considering the same to be a dynamic field since attackers are constantly devising new ways to attack equipment. Even if many new attacks are released daily, there will be more to come. Recent findings have demonstrated that the concept of a quantum computer has advanced from theory to a working technological system. Although they are still in the research and development stage, quantum computers are expected to surpass current supercomputers in performance. Furthermore, they note that programmers can soon fully implement quantum algorithms, which have the potential to crack all current forms of cryptography.

## 2.2 Fundamentals of Quantum Computing:

### 2.2.1 Quantum Computing:

The researcher [309] explains the concept of quantum computing and discusses the concepts of entanglement and superposition, which originate from quantum mechanics and are utilised in quantum computing, a novel form of processing that enables data to be processed in ways that conventional computers cannot. Classical bits can only be 0 or 1, while quantum bits, or qubits, can exist simultaneously in several states. Thanks to this capability, quantum computers can complete complex calculations more rapidly. This capability has much potential for addressing problems that conventional computers cannot, such as simulation and optimisation tasks. Quantum computing is based on manipulating measurable and controllable quantum states. In some instances, quantum computers can outperform traditional methods, as demonstrated by the development of quantum algorithms such as Grover's algorithm for searching through unsorted databases and Shor's algorithm for factoring large numbers. These algorithms dramatically accelerate computations by exploiting the inherent uncertainty and probabilistic nature of quantum physics. The finance sector sees quantum computing as a significant disruptive technology that can enhance risk analysis, financial modelling, and optimisation processes. The researcher further explains that through the simulation of complex financial systems and the simultaneous investigation of several alternatives, quantum computing offers a more advanced understanding of market dynamics and has the potential to improve financial decision-making.

### 2.2.2 Basic Building Blocks of Quantum Computing:

#### a. Qubit:

In quantum computing, a minuscule object, comparable in size to an electron or photon, is utilised by quantum computers to transfer stored digital information into the system. Encoding can be done only in one-bit information, either zero or one, through the two orthogonal states of a microorganism. A quantum bit, often known as a qubit, being a two-state system, is the primary unit of information in Quantum Computing, just like a binary bit in classical or traditional computing. New advances in energy, innovative materials, healthcare, environmental systems, and other fields are starting to adopt Quantum Computing.

#### b. Superposition:

The researcher [16] explains how superposition works; he mentions digital computers that reduce all the gathered data into a string of bits, after which logic gates are used in all calculations. So, if we consider the state of the digital computer, it is likely determined by its bits. So, if we want to discuss a laptop with n states, it can exist in any of the $2^n$ possible states. So, what is superposition? Now, quantum computers use bits that simultaneously signify a linear combination of zeros and ones rather than just one or zero.

#### c. Entanglement:

The activities of two different entities are linked by a phenomenon known as quantum entanglement. Researchers have discovered that changes to one entangled qubit instantly affect the other, a phenomenon necessary for quantum computing to function. It is possible to entangle pairs of qubits. It suggests that the two qubits reside in a single state. A foreseeable change in one qubit immediately affects the other under such circumstances, and Quantum algorithms, designed to handle complex problems, use this connection. Unlike conventional computers, where doubling bits doubles processing capacity, quantum computers experience an exponential increase in power with each added qubit.

**d. Interference:**

As discussed in the above paragraph about the superposition, there are probabilistic waves in a superposition state of entangled qubits. These stand for the probability of a system measurement's outcomes. When waves peak simultaneously, they may cancel each other out or build on top of each other at the intersection of peaks and troughs, which are the two instances of constructive and destructive interference, respectively. A quantum computer estimate requires first building a superposition of all possible computational states, where the user prepares the quantum circuit that applies interference to the constituent parts of the superposition. Many likely consequences are cancelled out by interference, while others are amplified, and we get the solutions to the computations as the detailed result.

## 2.3 Cryptography:

We are going to split the word cryptography, first considering the word prefix "crypt", which means "hidden," and the suffix "graph", which means "writing."The researcher [17] provides a comprehensive definition of the meaning and purpose of cryptography, explaining it as the study of secure, protected, and confidential transmission or communication techniques between two computers or parties. Researchers compare this mechanism with people's desire to communicate confidentially and privately with themselves. They take steps and measures to prevent third-party invasion and reading their messages. Researcher [18] explains in his article that nearly every encryption method used today is impacted by quantum theory. It uses mathematical algorithms and the algebraic structure of various protocols. It would not be wrong to say that it is the practice of securing information against possible adversaries who would obtain unauthorised access to confidential data if it were not for the mathematical guarantees of cryptographic systems.

### 2.3.1 Strategies for Cryptography:

In the modern era of computers, cryptography is frequently associated with converting plain text to ciphertext—text formatted in a way that only the intended authorised recipient can decode—a process known as encryption. Decryption is the process of converting encrypted text back to plain text. Cryptography and coding are ways to secure data, communications, or information, limiting access to only authorised people who can receive or understand this information or communication. The process involved prevents illegal access or invasion of data during the process. Considering the cryptographic technique, we can interpret it as safeguarding data, information, or communication using codes so that only approved and authorised recipients can access it. So, we can say that cryptography is a technique for preserving communications and information, ultimately preventing unauthorised access to confidential information. [19]Figure 1 briefs the encryption and decryption algorithm classification.

The researcher [20] clarifies that data transmission through a network requires security measures and strategies to prevent unauthorised persons from reading, altering, or intercepting the data illegally. During the 1960s, IBM dominated the IT market when it introduced its first-ever encryption method, which they named "Lucifer", and ultimately became the pioneer of the first Data Encryption Standard (DES). Cryptography is essential for protecting vast volumes of sensitive data, and its importance is only growing as our lives grow more

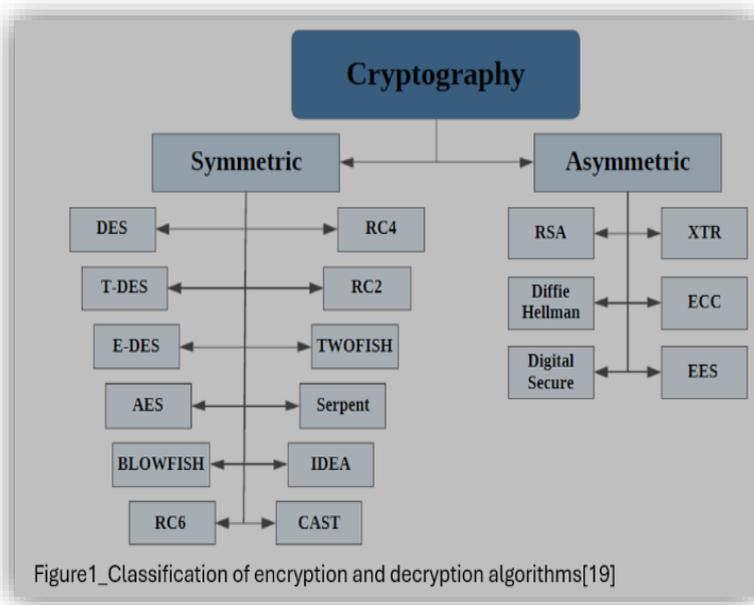

Figure1_Classification of encryption and decryption algorithms[19]

digital; it is not wrong to say that encryption plays a vital role in the internet world in numerous ways, so encryption is a necessary component due to large volume of sensitive data getting exchanged online every day.

## 2.4 Prior Year Research:

### 2.4.1 Cybersecurity in Accounting Research[2]:

The authors [2] have elaborated on the vital role of cybersecurity in safeguarding financial data in the accounting industry, which is examined in this study paper. With the development of digital technology, accounting businesses must implement robust cybersecurity measures due to the increasing potential for cyberattacks. This research has provided a comprehensive analysis of the role of cybersecurity in protecting financial data within the accounting profession. This study thoroughly examines how top accounting firms, such as Deloitte, PwC, and EY, have incorporated cutting-edge cybersecurity technologies, including advanced encryption, blockchain, and artificial intelligence, into their daily operations. For instance, Deloitte has implemented advanced encryption methods to protect sensitive financial data, while PwC has integrated blockchain technology for secure and transparent transactions. The study takes a qualitative approach, gathering information through case studies and interviews with accountants, IT specialists, and industry experts. It then applies the literature review methodology to analyse the state of cybersecurity in the accounting industry today, considering changes in cybersecurity protocols over time, new developments in the field, and potential directions for future research. This qualitative study employs interviews with accountants, IT specialists, industry experts, and case studies to obtain information. The efficacy of multi-layered cybersecurity solutions in enhancing data security, mitigating breaches, and strengthening fraud detection capabilities is underscored by significant findings. Nonetheless, issues such as exorbitant operating expenses and the rapid development of cyber threats persist.

## 2.4.2 Innovative accounting methodology for ensuring the interaction of economic and cybersecurity enterprises[21]:

In this article [20], the researcher aims to position accounting as a novel, multifaceted system that facilitates the interplay between economics and cybersecurity in businesses. In addressing the growing cyber threats that impact economic security, the study highlights the vital role that accounting plays in mitigating these risks. The primary objective of this article is to formulate and suggest an all-encompassing accounting approach that incorporates economic and cybersecurity measures at different tiers. Mitigating the financial impact of cyber threats requires modifying accounting standards, enhancing the accuracy of accounting data, and ensuring that reputation management and communication are effective. The gap identified by the authors is based on previous literature, where a separate study explored the intersection of economic security and cybersecurity. Research needs to be conducted to identify accounting as the bridge between them. This paper utilises various global datasets, including the Global Cybersecurity Index (2018), Global Innovation Index (2018), and World Digital Competitiveness Ranking (2018), to analyse the cybersecurity level of relationships between digital competitiveness across different countries and their economic security. The researcher's methodology in this article is an institutional approach to evaluating the response to cyber threats through modernising accounting practices. Future research, as identified by the researcher, emphasises the necessity of continuing studies into how accounting techniques can be constantly modified to counteract the ever-changing nature of cyber threats. The researcher suggests that one approach to this is to develop adaptable and responsive accounting procedures that can quickly adjust to emerging cyber threats. The researcher also noted the most significant losses, which are hardest to calculate, at the highest reputation level. Therefore, further research is necessary to accurately assess the financial impact of cyber threats.

## 2.4.3 Cybersecurity and Prevention in the Quantum Era[5]:

The author [5], in his article regarding the implications of quantum computing on cybersecurity, explained that among several industries, cybersecurity could be transformed through the progress and development of quantum computing at the 2nd International Conference for Innovation in Technology. Because quantum computers can execute a wide range of calculations far more quickly than classical computers, they may be able to crack many currently thought to be safe cryptography schemes. As a result, the development of quantum computers has sparked concerns regarding the security of current cybersecurity systems and the need to create new, quantum-resistant cryptography algorithms [22].

Organisations must remain knowledgeable and prepared to handle the evolving landscape of threats and defences in the quantum age, which presents both opportunities and challenges for cybersecurity. Cybersecurity prevents online theft, cyberattacks, and data loss on computer systems and linked networks. Combining technology, processes, and rules achieves systems security and prevents unauthorised data access. Cybersecurity risks come in various forms, including ransomware, phishing, malware, and denial-of-service attacks. Hackers, cybercriminals, and nation-state actors are the factors creating the danger. Quantum cryptanalysis, which analyses and potentially breaks classical cryptography systems by exploiting quantum states and measurements, is one possible application of quantum computers in cyberattacks. It could involve attacks against conventional cryptography protocols, quantum key distribution (QKD), and other quantum cryptographic protocols (IV).

Additionally, hackers may utilise quantum computers to carry out various types of cyberattacks, such as machine learning assaults, in which they train a quantum machine learning model to perform a specific function, like speech or image recognition. The author describes the quantum threat by considering two quantum machine learning algorithms: Shor's method and Grover's method. Shor's method is a quantum computing technique that factors large integers in polynomial time. It could undermine public-key encryption schemes, such as RSA, which rely on the idea that factoring large numbers is computationally impractical. Whereas "Grover's quantum search algorithm is a quantum algorithm that can search an unsorted database with N items in $O(\sqrt{N})$ time, which is faster than the $O(N)$ time required by classical algorithms."

### 2.4.4 Quantum computing for finance: Overview and prospects[4]:

The study [4] evaluates existing methods and explores future possibilities, discussing the potential applications of quantum computing in finance, including risk analysis, credit rating, and portfolio optimisation. It would have been interesting to examine how quantum technologies can be relevant to blockchain and cryptocurrencies. [23] Alternatively, to discuss quantum finance [24, 25], quantum money [26], the impact of quantum cryptography on the security of financial transactions [27, 28], and the applications of quantum simulators [29, 30] in finance. The study focuses on the application of quantum computing in risk analysis, credit rating, and portfolio optimisation in the banking industry. Quantum computation offers the possibility of highly efficient algorithms that could provide exponential speed-ups for some technologically significant tasks compared to classical information processing. [31].

The study claims that quantum annealers can optimise portfolios, uncover arbitrage opportunities, and evaluate credit. Quantum amplitude estimation offers a quantum mechanical approach to accelerating Monte Carlo sampling. The authors propose that researchers investigate the potential uses of malfunctioning quantum computers in finance before developing fault-tolerant quantum computing. The study explores deep learning in finance, quantum annealers, and quantum optimisation algorithms, and it makes recommendations for how quantum machine learning might enhance these techniques. The authors caution that several experimental breakthroughs are necessary before a universal quantum processor can be developed to surpass current supercomputers. They suggest that researchers explore the potential of quantum computing in finance.

Additionally, they propose that imperfect quantum computers could have exciting applications before fault-tolerant quantum computing is achieved. The authors claim that quantum computing has the potential to transform the financial industry. They encourage scholars to explore the potential applications of quantum computing in the financial sector. The authors assert that quantum computing can transform the field of finance. They urge researchers to explore the potential applications of quantum computing within the financial sector.

Table 1 summarises the consolidated and reviewed literature, where the research gaps are mapped to primary and secondary research questions (RQ1–RQ4) and article research propositions (P1–P4) to ensure relevant coverage of the study objectives. The table provides further direction for the study by demonstrating that the existing literature, to date, is insufficient in addressing numerous aspects related to quantum computing, cybersecurity, and accounting systems, including, but not limited to, vulnerabilities, quantum-resilient integration strategies, and institutional preparedness for financial quantum computing.

| Article Title | Research Question | Propositions | Identified Research Gap |
|---|---|---|---|
| **Cybersecurity and Prevention in the Quantum Era** | RQ1, RQ2 | Propositions 1 and 2 | This article identifies the critical gap in the transition from classical to quantum-secure encryption infrastructure in finance and accounting systems. It highlights vulnerabilities in current systems exposed by Shor's and Grover's quantum algorithms, showing that classical encryption (e.g., RSA, AES, ECC) will become obsolete. However, it lacks specific implementation frameworks or models for integrating post-quantum cryptography into organizational systems. |
| **Cybersecurity in Accounting Research (Haapamäki & Sihvonen, 2019)** | RQ1, RQ3 | Propositions 3 and 4 | The article synthesizes literature but reveals a lack of research addressing how accounting systems can be adapted to withstand quantum-era threats. Specifically, there is minimal attention given to integration challenges, protocol transitions, or organizational readiness for quantum-secure frameworks in accounting systems. |
| **Innovative Accounting Methodology of Ensuring the Interaction of Economic and Cybersecurity of Enterprises** | RQ3 | Propositions 3 and 4 | This work underscores the impact of cyber threats on accounting quality but lacks direct analysis of how quantum computing or quantum-safe measures can be incorporated into existing enterprise accounting methodologies. There's an evident absence of quantum-specific adaptation models or training frameworks for accounting personnel. |
| **Quantum Computing for Finance: Overview and Prospects** | RQ2, RQ3 | Propositions 1 and 4 | While offering technical insight into quantum optimization, machine learning, and amplitude estimation, this article does not bridge the gap into accounting-specific use cases. It lacks practical blueprints for integrating these algorithms into financial recordkeeping, audit trails, or encryption frameworks used in accounting systems. |

**Table 1** — Identified Research Gaps Mapped to Research Questions and Propositions

## 2.5    Limitations of the Existing Literature Reviews:

A key limitation of existing literature reviews is the failure to identify the intersection of quantum computing and cybersecurity in the context of accounting and finance. Currently, no reviews integrate these two domains and outline the unique implications, vulnerabilities, and opportunities for securing accounting and finance systems in the quantum era. It was pointed out that [32] lacks literature addressing the practicality of seamlessly integrating quantum computing technology with existing accounting and finance systems without disrupting ongoing operations. The transition from current systems to quantum-resistant alternatives has not been thoroughly planned or considered, and this process is expected to be quite complex. It was discussed [33] that, despite being considered a promising field of study for enhancing data security, the literature does not identify and evaluate the most promising quantum algorithms for securing financial data. It has left an important question unanswered: which quantum algorithms are the best suited to achieving cybersecurity in the financial sector? The author [34] points out that the lack of common standards for comparing and assessing quantum systems is a significant drawback, making it difficult to evaluate the potential and progress of quantum computing platforms unbiasedly. This lack of standardization also challenges the development of widely accepted quantum-resistant security measures. As noted in various studies, transitioning from current systems to post-quantum cryptography often requires substantial computational resources and a deep understanding of quantum technology. Additionally, the uncertain timeline for the arrival of large-scale quantum computers further complicates this transition. While theoretical advancements in quantum cryptography and quantum key distribution (QKD) are well-documented, few practical experiments and applications demonstrate how effectively these technologies can secure accounting and financial systems [34].

Financial data is heavily regulated (by laws such as GDPR for privacy, SOX for financial reporting integrity, etc.), and these regulations often mandate specific security controls. Regulators are beginning to recognise quantum threats; for example, government advisories in some countries have started recommending quantum-safe encryption for long-term sensitive data. [35]. However, clear guidelines for integrating quantum technologies in standard compliance frameworks are still emerging. It creates a knowledge gap that academic research can help fill by providing frameworks and evidence on the practical viability of quantum cybersecurity.

## 2.6 Cybersecurity Frameworks and Quantum Threat Landscape:

### 2.6.1 Classical Cybersecurity Frameworks in Accounting and Finance:

Financial systems rely on well-established cybersecurity frameworks to ensure the confidentiality, integrity, and availability (CIA triad) of digital information. Commonly adopted frameworks include:

- NIST Cybersecurity Framework (CSF): Widely used for risk-based cybersecurity management across sectors, including finance.
- ISO/IEC 27001: International standard for information security management systems.
- COBIT and COSO: Control-oriented frameworks are often used in internal auditing and financial reporting governance.

These frameworks integrate encryption protocols, identity access management, firewalls, and incident response procedures to protect financial data. [36]. However, these models largely depend on classical cryptographic algorithms such as RSA, Diffie-Hellman, and Elliptic Curve Cryptography (ECC). These rely on the computational difficulty of problems such as integer factorisation and discrete logarithms—assumptions that quantum computing can potentially break [3, 36].

### 2.6.2 Quantum Threat Landscape:

The advent of quantum computing fundamentally alters the security assumptions underpinning modern financial systems. Two primary quantum algorithms introduce new risks:

**a. Grover's concept:**

[37] Searching an extensive database is a problem of paramount importance, with applications in a wide range of fields. Since the Grover search algorithm offers a quadratic speedup in the number of database queries compared to classical computers, quantum algorithms can search a database significantly faster than classical computers. The Grover search algorithm is the optimal search quantum algorithm and may be used as a subroutine for other quantum algorithms. Searches with two qubits have been demonstrated on various platforms and proposed for others, but larger search spaces have only been demonstrated on a non-scalable NMR system. The Grover search algorithm has four segments: initialization, oracle, amplification, and measurement,

The initialisation stage creates a superposition over all states. The oracle stage marks the solution(s) by flipping the sign of the state's amplitude. The amplification stage reflects on the mean, thereby increasing the amplitude of the marked state. Finally, one measures the algorithm output. For a search database of size N, the single-shot probability of measuring the correct answer is maximised to near-unity by repeating the oracle and amplification stages $0((N)^{1/2})$ times. For a classical search algorithm, by comparison, the correct answer is found after an average of N/2 queries of the oracle. This quadratic speedup is a substantial advantage for large databases for quantum computers.

**b. Peter Shor's concept:**

The researcher [36] highlights that Peter Shor created the algorithm known as Shor's. He demonstrated that a quantum computer can factor in an integer N in polynomial time. According to [38], the precise real-time complexity is O(log N), where N represents the number of factors. The researcher [35] explains the significance of the algorithms, pointing out that they are significantly faster than the sub-exponentially fast, most efficient known classical factoring approach. Additionally, discrete logarithm issues have also been demonstrated to be solved by quantum computers. In the article [39], the author discusses the construction of what [40] Pitowsky refers to as a "clever superposition" of entangled qubits. Extracting a solution in a brief, or polynomial, amount of time, the factoring algorithm developed by the author [3] is considered the most significant quantum algorithm to date. It can factor large integers (numbers) into their constituent primes. Shor's method can potentially break RSA, the widely used public key encryption scheme on the Internet, in hours or days on a functional 100-bit quantum computer. When RSA public key encryption is scaled up to Quantum Kilo Bytes, it utterly fails. The researcher [39] explains that the Hidden Subgroup Problem (HSP) is a prominent class of quantum algorithms that includes Shor's algorithm. Variations of this issue have been identified that theoretically address the mathematical underpinnings of the three primary public-key encryption protocols currently in use on the Internet: RSA, ECDSA, and DSA [41]. The researcher [39] explains that Quantum technologies pose a significant threat to information security as they currently exist.

The author [36] notes that asymmetric or public key algorithms have become outdated and ineffective over time due to the advent of quantum computing, as it has been demonstrated that they can overcome the fundamental challenges of asymmetric cryptography. Further, he [36] explains that some network authentication protocols, digital certificates, VPN key exchanges, and all e-commerce systems can no longer use secure cryptography and secure Internet communication using TLS, commonly used to protect email, voice-over IP, and Web traffic, be possible. Anyone involved in cybersecurity must be aware of the issue and stay up to date on the developments leading up to a solution, even though these cybersecurity risks are not immediately apparent, as quantum computing is not yet a practical reality.

**c. Grover vs Peter Shor's Algorithm and Accounting and Finance:**

Each poses distinct challenges to the classical cryptographic infrastructure underpinning accounting and financial systems. Among these, Shor's algorithm is particularly critical for the finance and accounting sector, as it directly compromises the integrity of public key cryptography (e.g., RSA, ECC)—widely employed for digital signatures, secure financial reporting, and audit trail authentication [42]. The author [33] pointed out that modern cryptographic systems, such as RSA, rely on the difficulty of factoring large numbers as a foundation for their security. If, as highlighted by the author [33], quantum computers advance to the point where they can effectively use Shor's algorithm, sensitive financial data—most of which is protected by these cryptographic techniques—could be at risk. The researcher [43] discusses RSA encryption, which is utilized in various cloud-based accounting programs. If quantum computers capable of running Shor's algorithm become available, attackers could exploit these systems, leading to potential data theft or manipulation. It could significantly undermine customer trust and disrupt company operations. The researchers [44, 45] highlighted the concerns regarding digital signatures, secure communications, and blockchain-based systems, which are commonly employed in financial reporting and auditing, as they are particularly vulnerable to Shor's algorithm, which can break RSA encryption in polynomial time. This vulnerability underscores the need to develop quantum-resistant cryptographic systems to ensure the security of financial transactions and data. The author [33] has also noted that Shor's algorithm has significant implications for the accounting and finance industry, extending beyond the need for enhanced cybersecurity. He further explains that companies must invest in the research and development of new cryptographic standards that can withstand quantum attacks in order to prepare for the potential impact of quantum computing. Building on his explanation above, he emphasised the importance of protecting the integrity and confidentiality of financial data, which is crucial, as prospective breaches could compromise this information. The researcher [46] notes that transaction records could lose immutability in the event of a quantum breach. The cryptographic hash functions used by blockchain technology, which are commonly employed to ensure data integrity, may be vulnerable to quantum attacks. If these cryptographic protections are compromised, financial records could be altered without authorization, undermining trust in the reliability and accuracy of financial data. Therefore, the accounting and banking sectors need to develop quantum-safe procedures and technologies urgently [33], and this necessity is underscored by the implications of Shor's algorithm as the industry adjusts to the emerging quantum landscape. In his article, the researcher [44] also emphasises developing encryption techniques that can withstand quantum attacks. Quantum Key Distribution (QKD) and lattice-based cryptography are two promising approaches that are gaining popularity.

## 3.0 Conceptual Framework and Research Propositions:

### 3.1 Justification for Theoretical Lens:

This study employs a multifaceted theoretical framework to examine the adoption of quantum-secure cybersecurity measures within accounting and finance systems. The selection of theories is driven by the necessity to approach technology adoption from individual, organisational, and domain-specific perspectives. At the personal level, the study references the [47]Thompson et al.'s (1991) Model of Personal Computing Utilisation as a foundational framework for assessing user adoption of digital innovations. Six constructs adapted from information systems and contingency theory examine structural and cultural readiness within institutions, including task interdependence, technology, environment, structure, goals, and people. [48, 49]Additionally, in light of quantum computing's unique characteristics, the study introduces four quantum-specific constructs, two of which are original to this research: Quantum Resistance of Accounting Algorithms and Quantum Key Distribution (QKD) Integration. This layered framework offers theoretical comprehensiveness by linking established information systems adoption models with emerging quantum security considerations pertinent to financial systems.

## 3.2 Integrated Conceptual Framework of Finance and Accounting System Leveraging Quantum Computing-based Cybersecurity Strategies:

The integrated conceptual framework encompasses three dimensions: (1) user-level constructs from [47]Thompson et al. (1991), (2) organisational-level constructs based on contingency theory, and (3) quantum-specific cybersecurity constructs. This design aims to comprehensively understand how quantum tools impact cybersecurity adoption in accounting settings. The framework has been refined through a systematic literature review following the PSALSAR methodology and visually depicts the dynamic relationships among these constructs. Each proposition in this study is assessed based on factors that ensure thematic alignment and theoretical relevance, grounded in theoretical precedent and empirical necessity.

### 3.2.1 User-Level Constructs: Technology Adoption Models:

These constructs are grounded in [47]Thompson et al.'s (1991) Model of Personal Computing Utilisation, which has been widely validated in information systems (IS) research to understand individual attitudes and behaviour toward emerging technologies [50]. This adaptation explores how quantum computing can enhance data security, protect sensitive financial information, and improve system integrity and confidentiality. This model comprises six key factors, but we focus on more relevant perceptions: job fit, complexity, perceived consequences, and the effect on use. These constructs were later extended in models such as the Unified Theory of Acceptance and Use of Technology (UTAUT) [50] and TAM 3 [51], which also prioritise user-centric adoption factors. In the context of accounting, user acceptance among auditors, accountants, and finance controllers is a critical prerequisite for the successful implementation of any cybersecurity framework. These factors are crucial in understanding how an organisation's operations work and how its processes integrate AI into its system. The conceptual model for personal computing utilisation proposed by [47], of which we have mentioned the initial four factors, now we discuss the last two new factors integrated into this conceptual model: the first is the relationship, and the next is the culture. After integrating quantum computing into the cybersecurity systems of accounting and finance, these factors are critical to analysing the organisational impact, concerns, and shifts due to cultural changes.

#### a. Job Fit/Job Performance:

Job fit, for instance, involves assessing how quantum computing can enhance job performance and decision-making processes within an organisation. In his article, the researcher [47] defined job fit as a scale that measures an organisation's acceptance of introducing new technology, thereby enhancing job performance. We consider the same application for our quantum computing research in cybersecurity, specifically in the fields of accounting and finance. We need to consider quantum algorithms to analyse the job-fit factor. The most critical factor of accounting is financial information, and these algorithms have the potential to improve their precision and efficiency remarkably. Accountants can apply

Quantum computing tools to optimise complex financial models, reduce errors, and streamline operations, thus improving job performance and lowering operational costs. Accountants and auditors may find these quantum tools more aligned with their job needs, mainly where big data is involved, resulting in enhanced productivity and effectiveness.

**b. Complexity:**

Complexity, as the word describes, encompasses the perceived difficulty in understanding, applying, and managing innovations related to quantum computing, particularly when applied to big data and diversified information, much like the application of new technology, as explained by the researcher [47]. Quantum computing must perform complex computations due to the quantum algorithms we have discussed previously; therefore, it would not be incorrect to say that they are complex and require sophisticated mathematical foundations.

**c. Perceived Consequences:**

According to [47], Perceived consequences involve the belief that utilising a particular technology leads to beneficial outcomes, such as increased job satisfaction or efficiency. In the context of quantum computing, the perceived consequences in accounting cybersecurity include enhanced protection of financial data against future quantum attacks, reduced risk of breaches, and compliance with emerging regulatory requirements. Implementing quantum-resistant algorithms ensures long-term security, while quantum key distribution (QKD) provides secure communication channels, thereby enhancing the overall perceived benefits of adopting quantum technologies in accounting systems. Perceived consequences would focus on the potential outcomes and benefits of integrating quantum computing into operations management, such as improved efficiency and decision-making speed.

**d. Affect Towards Use:**

The researcher's [47] conceptual model factor, "Affects towards use," measures emotional responses to technology, including trust, satisfaction, and frustration. The integration of quantum computing into cybersecurity for accounting is likely to generate positive emotions among users due to its potential to provide unprecedented levels of security. For instance, quantum-resistant algorithms can offer a robust defence against even the most sophisticated cyber threats, instilling a sense of trust and reliability in the system.

**e. Relationship/Social Factor:**

As discussed by [51], the relationship factor involves the interactions among rational decision-makers, as analysed using game theory. In accounting cybersecurity, this could model the relationships between financial institutions, auditors, and regulators, where the potential benefits of quantum security measures influence each player's strategy.

**f. Culture:**

Culture encompasses an organisation's collective values, beliefs, and traditions that influence the adoption of technology. A culture prioritising cybersecurity and innovation is more receptive to integrating quantum computing technologies. Organisations that foster a security-conscious culture are more likely to embrace quantum-resistant algorithms, recognising their importance in protecting financial data against emerging threats. Promoting a culture of quantum awareness and continuous learning is crucial in ensuring that the full benefits of these advanced technologies are realised.

**3.2.2  Organisational Constructs:** *IS/Contingency Theory Foundations:*

To expand from individual acceptance to organisational readiness, the model incorporates six constructs frequently applied in IS security adoption literature and Contingency Theory [49]:

**a. Task Interdependence:**

Quantum computing allows for faster and more accurate processing of complex tasks in accounting and finance. For instance, quantum algorithms can significantly accelerate Monte Carlo simulations, which are essential for pricing derivatives. This process relies heavily on large amounts of interconnected financial data. [52, 53].

**b. Technology:**

Quantum algorithms, particularly those for amplitude estimation and quantum walks, can solve critical financial problems that are intractable in classical computing, including enhanced financial modelling, risk assessment, and simulation capabilities [52].

**c. Environment:**

This quick and efficient data analysis can also be applied in the financial world to enhance risk and uncertainty prediction, providing valuable insights into market fluctuations. Modelling financial uncertainties as quantum states leads to a more nuanced understanding of market dynamics. (Placing quantum finance on the agenda may also help to convince those wary of quantum technologies that there is more to gain from the metaphysical aspects of the quantum than they might otherwise think) [52].

**d. Structure:**

Financial institutions' organisational processes likely need to change and incorporate new components, including new data pipelines and an end-to-end digital twin, to assess how these quantum models are performed, enabling more efficient and secure banking services [54].

**e. Goals:**

The driving force behind quantum computing in finance is the promise of enhanced data security and computational efficiency. Quantum computing can enhance financial data security through quantum-resistant cryptographic techniques. Additionally, [54, 55] can enhance other computational tasks, such as credit risk computations and economic forecasts.

**f. People:**

Humans successfully oversee the use of quantum computing harnesses. Quantum technologies and cybersecurity must be applied and taught to finance professionals so they can use them correctly and effectively. Researchers, financial institutions, and regulators must exchange ideas and expertise closely to ensure the appropriate deployment of emerging quantum technologies.

**3.2.3 Quantum-Specific Constructs: Emergent from Empirical Synthesis:**

The third layer, quantum-specific cybersecurity constructs, was first developed based on identified gaps in existing IS adoption models (which do not cover quantum cryptography or QKD), and secondly, through a Systematic literature review (via PSALSAR) of Q1 cybersecurity and quantum computing journals. Another basis is the empirical propositions grounded in studies like [56].

This dimension is novel and essential for this article because traditional IS frameworks do not account for post-quantum security implications, directly impacting encryption durability, audit trails, and financial compliance systems. These constructs include:

   a. **Quantum Resistance of Accounting Algorithms:**

The researcher [57] explains that quantum resistance refers to the robustness of cryptographic algorithms against quantum attacks, also known as post-quantum cryptography. As traditional encryption methods become less effective, quantum computers gain more power in accounting systems and may be vulnerable to being broken by quantum algorithms, such as Shor's. Integrating quantum-resistant algorithms, such as lattice-based or hash-based cryptography, into accounting systems is crucial for maintaining the security of financial data. This factor emphasizes the importance of forward-looking cybersecurity strategies that can withstand the potential threats posed by quantum computing.

   b. **Quantum Key Distribution (QKD) Integration:**

The researcher [58] is considering QKD technology, which utilises quantum mechanics to securely distribute cryptographic keys between parties. This method is unique in its ability to detect eavesdropping attempts, making it a highly secure means of communication. In accounting, integrating QKD into cybersecurity protocols can ensure that cryptographic keys protecting financial information remain secure, even in the face of quantum-enabled threats. This factor highlights the transformative potential of quantum technologies in enhancing the security infrastructure of accounting systems, making them more resilient against future cyber threats.

   c. **Organisational quantum readiness:**

[59] Describes the degree to which institutions possess the infrastructure, digital literacy, and strategy alignment to pilot or scale quantum technologies [60]. Preparedness of the institution in terms of digital infrastructure, leadership support, and financial capacity to pilot or adopt quantum systems.

   d. **Stakeholder interdependence:**

[61]It highlights the necessity of coordination across finance teams, IT security units, vendors, and auditors to ensure policy compliance and the secure deployment of quantum tools. [62]Accounting cybersecurity depends on interaction among internal and external actors (CFOs, auditors, compliance, regulators, and IT vendors).

The modelling framework in this study incorporates user-level factors from [47]Thompson et al.(1991), organisational-level constructs from IS adoption literature, and newly identified quantum-specific constructs. These 16 dimensions provide a framework for assessing the readiness, alignment, and impact of quantum-secure cybersecurity on accounting and finance systems. The model, as illustrated in Figure 2, shows how individual beliefs influence adoption and how institutional goals, team interdependencies, and stakeholder interactions facilitate or hinder technological transformation. Quantum-specific constructs (QKD, post-quantum algorithm resistance) introduce new urgency to this sector's cybersecurity strategy.

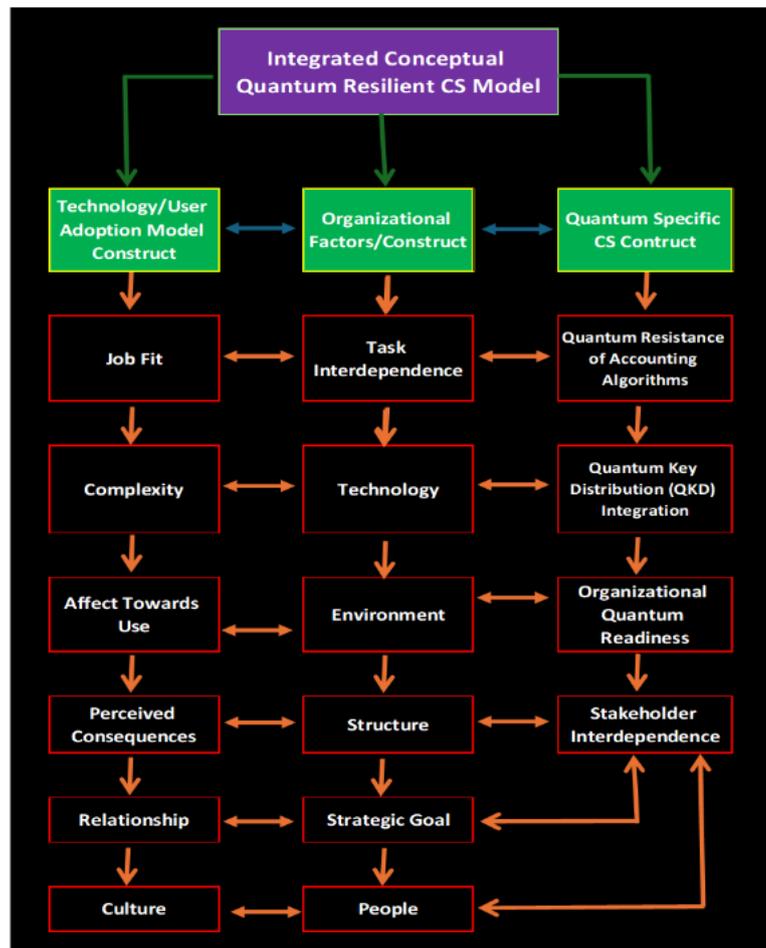

Figure 2: Integrated Conceptual QC Model

### 3.3 Mapping the Integrated Conceptual Framework Factors to Research Objectives and Questions:

The article's research questions and propositions are grounded in the integrated framework outlined above. Research Question 1, which investigates the relative security benefits of quantum computing over classical systems, draws upon the constructions of Quantum Resistance, Job Fit, and Technological Goals. Research Question 2, focused on unauthorised access prevention, is analysed through the lens of QKD Integration, Task Interdependence, and Structure. Research Question 3 addresses organisational adaptation and cultural shifts informed by the Concepts of Complexity, People, and Culture. Finally, Research Question 4 examines adoption barriers and long-term values, evaluating them using the Environment, Digital Readiness, and Perceived Consequences constructs.

### 3.4 Research Propositions:

Based on the integrated conceptual framework presented in this study, four research propositions have been formulated to explore the implications of quantum computing on cybersecurity within accounting and finance systems. These propositions are grounded in theoretical constructs and thematic insights from the systematic literature review and conceptual synthesis. Each proposition reflects a distinct but interrelated domain of inquiry within the study's overarching objective.

#### 3.4.1 Proposition 1:

Quantum computing provides stronger encryption and security than classical systems in the fields of accounting and finance. This proposition examines the hypothesis that quantum technologies, particularly lattice-based post-quantum cryptography, can mitigate vulnerabilities inherent in traditional encryption protocols, such as RSA and ECC, which are widely used in financial record-keeping and audit trails.

### 3.4.2 Proposition 2:

Quantum-secure cybersecurity frameworks will prevent unauthorised access and protect sensitive financial and accounting data from cyberattacks. This includes using Quantum Key Distribution (QKD) to establish mathematically secure, tamper-evident communication between stakeholders in accounting ecosystems.

### 3.4.3 Proposition 3:

Adopting quantum-secure cybersecurity frameworks will necessitate significant adjustments to security protocols, professional training, and organisational culture. This proposition addresses the institutional and human capital challenges of transitioning to quantum-secure infrastructures.

### 3.4.4 Proposition 4:

Challenges such as high costs, the need for specialised skills, and the complexity of quantum computing will initially hinder its widespread adoption in the accounting sector. However, the enhanced security benefits will eventually outweigh these challenges, especially as regulatory environments evolve and digital readiness improves.

## 4.0 Research Methodology:

### 4.1 Introduction:

#### 4.1.1 Systematic review:

The author [63] explained that a systematic review gathers all empirical data that satisfies predetermined qualifying requirements to address a particular research topic. It employs precise, methodical techniques to reduce bias, resulting in trustworthy results from which judgements and conclusions can be reached [64, 65]. According to the researcher [66], a systematic review is a straightforward and reliable technique for synthesising scientific data to address a specific research question. Their objective is to consider all published data currently available on the subject and assess the quality of the material. The researcher [67] emphasises the review protocol, which must be created, pertinent databases must be found, suitable search words must be developed, inclusion and exclusion criteria must be established, data must be analysed, and conclusions must be synthesised. These are just a few crucial elements in the process [68].

#### 4.1.2 Meta-Analysis:

The researcher [69] explains that the meta-analysis integrates and condenses the findings of included studies through statistical methods. It is present in many, but not all, systematic reviews. By combining data from all the relevant studies, meta-analyses can yield more accurate estimates of the effects of the research data than individual studies included in a review.

In their research article [69], twenty-nine review writers, methodologists, doctors, medical editors, and consumers collaborated to develop the PRISMA Statement. The author [70] in the article provides an evidence-based minimum set of guidelines; the PRISMA Statement and its expansions seek to promote

Page **21** of **44**

complete and transparent reporting of social return [71]. Other researchers [72] refer to the Preferred Reporting Items for Systematic Reviews and Meta-Analyses (PRISMA) Statement, the most widely used reporting standard for systematic reviews, which covers the literature search component [69].

### 4.2 Research Methodology adopted:

The researchers [73] emphasise that the essential elements of systematic reviews include the use of exact and repeatable methods for selecting primary research publications, as well as the critical assessment and synthesis of studies that meet the eligibility criteria [74-76]. The research methodology to be adopted for performing this research is the same as explained in the article by Grover [50]. Grover chose to use the systematic literature review to answer his research questions. The article employs the same methodology to address the research questions regarding accounting-related cybersecurity based on quantum computing. The researcher's [77] conclusions indicate that a systematic literature review may evaluate and synthesise all pertinent research on a particular research question, topic area, or phenomena of interest (SLR) [78]. This study employs a qualitative, interpretive paradigm to understand the intersection of quantum cybersecurity and accounting information systems. A conceptual and theoretical inquiry was chosen to explore the emerging field where empirical data is limited, but theoretical foundations can still be rigorously constructed. The purpose of an SLR is to provide a reliable way to gather objective, reasonable, and unambiguous data on a study topic [79]. The Kitchenham and Charters [80] systematic review adheres to principles. According to them, a predetermined, rigorously adhered-to protocol lowers researcher bias and improves rigour and repeatability. Given the novelty of quantum computing applications in finance, this study employs a systematic literature review (SLR) as the primary method to identify, analyse, and synthesise existing scholarly knowledge.

#### 4.2.1 PSALSAR Research Methodology:

This systematic literature review in quantum accounting is being conducted using PSALSAR. This methodology was created in 2020, as detailed in this article [81]. The researcher [82], in his scholarly article, mentioned that the framework for Search, Appraisal, Synthesis, and Analysis (SALSA) is a methodology that ensures methodological precision, systematisation, exhaustiveness, and repeatability in determining the search protocols that the systematic literature review (SLR) should adhere to. According to Grant and Booth [83], two steps are added to the standard SALSA technique to create the new PSALSAR method, which employs six phases to conduct systematic literature reviews. Research Protocol (P) and Reporting Results (R) are the latest phases. The PSALSAR methodology is straightforward, portable, and repeatable. The researchers [82] further explained that this methodological technique was employed in most scientific works [82-85] to mitigate the risk of publication bias and enhance the work's acceptability. The Preferred Reporting Items for Systematic Reviews and Meta-Analyses [84] and the Search, Appraisal, Synthesis, and

Analysis (SAS) framework [83]; most review articles thus adhered to the framework in their literature

Table : 2 PSALSAR framework for systematic literature reviews.

| Framework | Steps | Outcome | Methods |
|---|---|---|---|
| PSALSAR Framework | Protocol | Defined study scope | Research Question, PICOC framework to be considered |
| | Search | Define the search strategy<br>Search studies | Searching strings<br>Search databases |
| | Appraisal | Selecting studies<br>Quality assessment of studies | Defining inclusion and exclusion criteria<br>Quality criteria |
| | Synthesis | Extract data<br>Categorize the data | Extraction template<br>Categorize the data on the iterative definition and ready it for further analysis work |
| | Analysis | Data analysis<br>Result and discussion<br>Conclusion | Quantitative categories, description, and narrative analysis of the organized data<br>Based on the analysis, show the trends, identify gap and result comparison<br>Deriving conclusion and recommendation |
| | Report | Report writing<br>Journal article production | PRISMA methodology<br>Summarizing the report result for the larger public |

Source: Modified from (Del Amo et al., 2018, Mengist et al., 2020a)

searches.

Table 2 outlines the procedures employed in this literature review on cybersecurity for accounting. This systematic study aims to provide an extensive overview of published, peer-reviewed works related to quantum accounting. The review framework is based on the PSALSAR model, which provides a structured process across planning, searching, appraisal, synthesis, analysis, and reporting stages. Using the Conceptual Framework, we respond to the abovementioned research questions through this process. We follow all three steps, as outlined in Table 2.

*Step 1_Protocol:*

Table : 3 SLR research scope based on the application of the PICOC framework to the determined objectives

| Concept | Definition according to Booth et al. [12] | SLR application |
|---|---|---|
| Population | The research work dealing with cybersecurity for accounting based on Quantum Computing. | Scientific research work on dealing with the Cybersecurity for accounting based on Quantum Computing, Research studies, academic papers, and technical reports focused on the application of quantum computing in enhancing cybersecurity in accounting. |
| Intervention | Existing techniques utilized to address the problem identified. | Incorporating Quantum Computing in the accounting field. Current cybersecurity techniques in accounting, such as encryption methods, and how they can be enhanced or replaced by quantum computing. |
| Comparison | Performing SWOT analysis | Identifying the strengths, weaknesses, opportunities and threats of adopting the Quantum Computing in accounting for Cybersecurity. Comparing quantum computing methods with classical computing techniques in terms of effectiveness, scalability, and practicality for cybersecurity in accounting. |
| Outcome(s) | Measure to assess the knowledge and gaps mentioned in the selected publications related to Quantum based accounting. | Identification of gaps in current research regarding the integration of quantum computing in cybersecurity for accounting and proposing areas for future study. |
| Context | The particular settings or areas of the population. | Trends of the Cybersecurity for accounting based on Quantum Computing research, their challenges, gaps. The specific context within accounting (e.g., financial data protection, auditing, fraud detection) where quantum computing could be applied to improve cybersecurity. |

Source: Modified from [84, 81]

As shown in Table 3, the PICOC tool was used to complete the search procedure.

*Step 2:_Search:*

This stage includes the delivery and search strategy. To gather pertinent documents, the search strategy helps define a suitable search string and identify relevant databases. [85]. The researchers further explained in their article that using the search phrase to access the chosen databases and gather numerous related literature papers is part of the search delivery stage. The search used two reputable and well-known databases: Web of Science and Scopus.

- **For Web of Science**, the search was done using the following:

*QUANTUM (All Fields) and COMPUTING (All Fields) and ACCOUNTING (All Fields) and FINANCE (All Fields) or CYBERSECURITY (All Fields) and INFORMATION (All*

*Fields)* and Enriched Cited References and **2024** or **2023** or **2022** or **2021** or **2020** or **2019** or **2018** or **2017** or **2016** or **2015** (Publication Years).

- **For Scopus**, the search was done using the following:

( TITLE-ABS-KEY ( quantum ) AND TITLE-ABS-KEY ( computing ) AND TITLE-ABS-KEY ( accounting ) OR TITLE-ABS-KEY ( finance ) OR TITLE-ABS-KEY ( financial AND reporting ) OR TITLE-ABS-KEY ( cybersecurity ) ) AND PUBYEAR > 2014 AND PUBYEAR < 2025

The words were chosen to encompass as much scientific literature as possible on quantum computing, cybersecurity, accounting, and finance. Initially, the years of publication for both Web of Science and Scopus were considered for the last ten years, from 2014.

*Step 3:_ Appraisal:*

The researcher [82] is considering the appraisal phase; the chosen articles were assessed following the goals of the review process. Screening the chosen literature to find pertinent papers for the review process was part of the research selection process. The two main phases are the quality assessment and the selection of studies based on inclusion criteria. The studies were selected by applying the inclusion and exclusion criteria, as shown in Table 4. The papers underwent thorough screening to select the relevant articles for the review as part of the quality assessment process. Initially, the titles and abstracts of the detected records

**Table 4 SLR study selection of literature using inclusion and exclusion criteria.**

| Criteria | Decision |
|---|---|
| When the keywords exist in title, keywords or abstract section of the paper. | Inclusion |
| The paper published in a scientific peer-reviewed journal or article. | Inclusion |
| The paper should be written in the English language | Inclusion |
| Papars contain concepts relevant to objective of the research. | Inclusion |
| Provides information on CS or quantum theory related to accounting. | Inclusion |
| Papers that are duplicated within the search documents | Exclusion |
| Papers that are not accessible, review papers and meta-data | Exclusion |
| Papers that are not primary/original research | Exclusion |
| Papers written in languages other than English. | Exclusion |
| Grey literature, chapters, editorials,notes, letters, new items, surveys | Exclusion |

**Source: Modified from [81]**

were reviewed, and papers that did not meet the inclusion criteria were excluded. Subsequently, the full texts of the remaining records were assessed for eligibility.

*Step 4:_Synthesis:*

Furthermore, the researcher [81] is considering the synthesis process, which involves extracting and classifying pertinent data from selected papers to develop information and conclusions. Extracting data from the selected articles entails locating and obtaining relevant information.

Two research databases were chosen: Web of Science and Scopus. From Web of Science, 2,334 documents and 672 from Scopus, a total of 3,006 documents were selected through the keywords described in step 2. Papers were extracted for further review. 57 duplicated papers were removed through website screening. Twelve documents of a language other than English were removed. Furthermore, 353 published journals were excluded according to the inclusion/exclusion criteria, and the remaining articles were retained for further review. The next step we took was to limit our articles to the last three years for publication; articles from the prior year were excluded, resulting in 212 completed articles.

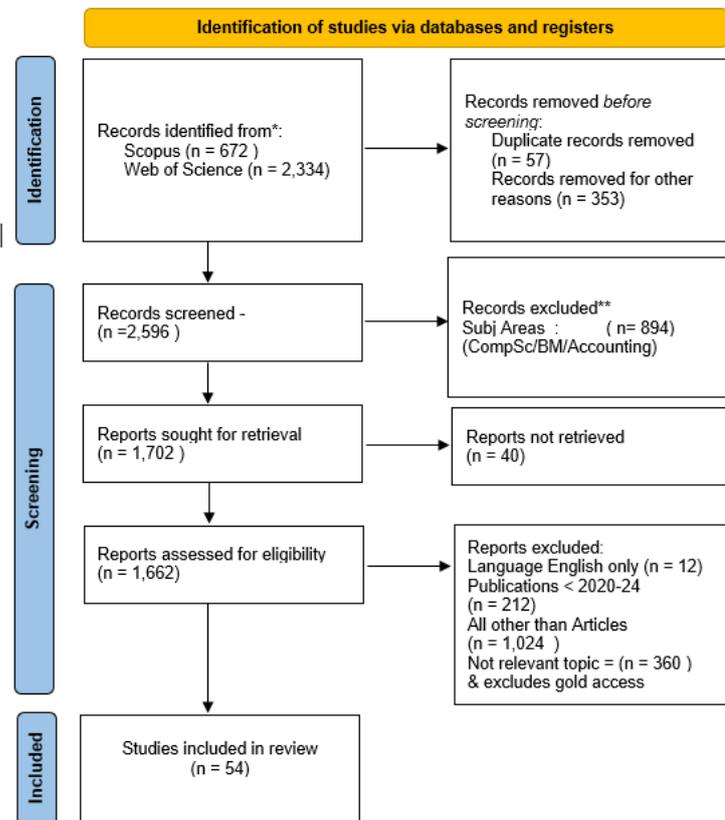

[62] Flow Diagram_1 _Prisma Flow Diagram:

Publications that were articles were considered, eliminating the remaining publications by 1,024.

We also considered keeping articles with research areas in Computer Science and Business while excluding other research areas, which resulted in 894 more articles being dropped from the final list. The total number of articles remaining is 454, which can now be assessed by screening the title. Titles without "Quantum" plus those not able to be retrieved will be excluded from the selection, resulting in 54 articles remaining as selected. It has been detailed as per Flow Diagram No.1.

*Step 5 Analysis*:

The researcher's consideration [81] is the analytical step, which includes assessing the combined data, identifying significant material, and drawing conclusions from the chosen articles. The researcher's [85] study goal, the analyst's assessments, and the researcher's understanding of the research's level all influence analysis and result reporting. The Microsoft Excel spreadsheet was used for data extraction and analysis.

## 5.0 Findings & Discussions of Implications for Cybersecurity

This study posits that quantum computing holds transformative potential for the accounting and finance sector, particularly when analysed through our adapted version of Thompson's Conceptual Model [46]. The model's core dimensions—task interdependence, technological advancement, environmental adaptability, and organisational structure—are all positively influenced by emerging quantum technologies. Rather than treating quantum computing as a mere technological novelty, it should be viewed as a strategic enabler capable of reshaping decision-making frameworks, enhancing process efficiency, and securing financial data systems. When strategically implemented, quantum tools can improve the alignment between organisational operations and long-term financial objectives. As evidenced in the following analysis of the four propositions, quantum computing supports the reconfiguration of internal processes, enhances adaptability to regulatory and market environments, and facilitates strategic outcomes in accounting and finance. This positions quantum computing as a critical lever for advancing resilience, precision, and agility in financial decision-making and information systems.

### 5.1 Results and Findings – Testing of Propositions:

We adopt an integrated conceptual model. The hybrid theoretical framework synthesises multiple scholarly foundations to examine the four core propositions outlined in this study: Quantum Computing and Cybersecurity in Accounting and Finance: Current and Future Challenges and Opportunities for Securing Accounting and Finance Systems. Each proposition is evaluated through a targeted combination of these variables, selected based on thematic alignment and conceptual relevance. This multidimensional approach enables a rigorous and context-specific analysis of how quantum computing is poised to impact cybersecurity practices within accounting and finance systems.

#### 5.1.1 Proposition 1:

**Quantum computing offers significant advancements in encryption methods, enabling more effective prevention of cyberattacks on accounting systems compared to classical computing.**

   a. **Job Fit/Job Performance:** Articles by [33, 86] emphasise that quantum-secure encryption tools—such as lattice-based post-quantum cryptography (PQC) and Quantum Key Distribution (QKD)—align well with key accounting functions, including audit trail protection, ledger verification, and secure interdepartmental financial reporting. Job relevance is established in scenarios like post-quantum audit verification and secure cloud bookkeeping.

   b. **Complexity:** Over time, quantum computing could bring quantum-resistant encryption into the mainstream, allowing accountants to learn how to apply it, and new tools will make it easier to use. Hybrid models are emerging, combining classical encryption for ease of use with quantum protocols for secure key distribution, suggesting increasing accessibility [87]. [88] It has been noted that quantum encryption introduces higher mathematical complexity and infrastructure burden. While secure, implementation requires specialised skills, posing adoption barriers in mid-size accounting firms.

   c. **Perceived Consequences:** More crucial is the role of quantum computing in providing much more effective encryption regimes, such as Shor's algorithm, which can rapidly crack any existing classical encryption scheme. It enhances the effectiveness of quantum encryption in securing accounting systems. Quantum computing enables cryptographic protocols (e.g., QKD) that are provably secure and immune to computational attacks, making it a superior choice for accounting data protection [44].

   d. **Attitude Toward Use:** However, as organisations become more aware of the threat of quantum-capable cyberattacks, positive attitudes towards quantum-based encryption emerge. Increasing awareness of quantum threats fosters proactive attitudes towards adopting quantum cryptographic methods [89].

e.  **Behavioural Intention to Use:** Quantum encryption technologies are becoming firmly entrenched in accounting firms as the threat of future quantum hacking turns from fiction to fact. Institutional momentum is building toward post-quantum transition, with initiatives already launched in the financial and defence sectors rising [90].

f.  **Relationship/Social Factors:** [32] Highlight regulatory pressure (e.g., NIST PQC call) driving social and professional urgency to adopt quantum-ready security. Peer pressure within the finance industry is also on the rise.

g.  **Technology Characteristics:** Quantum Key Distribution (QKD) offers unbreakable encryption secured by the laws of physics, unlike classical encryption, which relies on computational difficulty [58]. Shor's algorithm can break RSA and ECC, underscoring the need to transition to post-quantum and quantum-native encryption protocols [87]. Several studies, including [4], demonstrate that technologies such as CRYSTALS-Kyber and quantum entropy-based encryption outperform classical methods in resisting decryption. These represent critical innovations for accounting systems vulnerable to replay and ledger modification attacks.

h.  **Organisational Structure:** Governments, academic research institutions, and software companies collaborating to provide resources and facilitate the adoption of technology would best serve the development of quantum encryption. Global collaboration across academia, governments, and tech companies is accelerating the research and deployment of quantum-secure systems [91]. Accounting firms must revise their protocols and approval chains to support decentralised, quantum-resistant verification systems. [92].

i.  **People:** Not only do accountants recognise cybersecurity's importance to confidential accounting data, but firms that innovate by following technology trends also turn to quantum-resistant encryption. Adoption is driven by a sense of urgency across institutions aware of the quantum threat, with early adopters (e.g., financial institutions) expected to set industry norms [44]. A lack of post-quantum cybersecurity skills remains a bottleneck. Most accounting personnel lack awareness of PQC constructs, such as NTRUEncrypt or CRYSTALS-Kyber [93].

j.  **Strategic Goals:** Across studies, organisational goals include fraud prevention, data sovereignty, and reduced forensic costs. Quantum-secure encryption aligns better with these goals than classical encryption due to its provable security guarantees under quantum threat models [32].

k.  **Quantum Resistance of Accounting Algorithms:** This factor emphasises the need to develop and subsequently use quantum-resistant algorithms designed specifically for accounting systems. These algorithms shield against potential quantum cyberattacks, as no quantum computer surpasses them and falls prey to a malicious agent. Multiple studies confirm that classical algorithms are vulnerable to quantum attacks, necessitating the development of quantum-resistant upgrades for financial systems [44]. [94] [95] confirms that quantum computers can break RSA, ECC, and SHA-256. Studies promote lattice-based schemes (e.g., Kyber, Dilithium) as replacements. There is a strong consensus that quantum-resilient cryptography must replace classical algorithms in financial systems.

**l. Quantum Key Distribution (QKD) Integration:** QKD offers an unbreakable, secure, essential exchange that is also secure against quantum computers. It considerably increases the security of accounting systems. QKD ensures secure key exchange and prevents eavesdropping, even from quantum-enabled attackers [58]. Quantum computing brings new kinds of robustness to the efficacy of encryption against cyberattacks. Quantum-resistant algorithms are designed to resist both classical and quantum computers when executing cryptanalysis. Undeniably, [96] quantum computing becomes more promising with the effort being dedicated by world governments and the comprehensiveness of quantum-resistant cryptographic solutions. We do not envision the quantum transformation outpacing the development of quantum-resistant cryptography, but the end of our current era is likely to occur sooner rather than later. [33] QKD ensures secure key exchange over quantum channels, even in adversarial settings. QKD supports end-to-end confidentiality for interbank reconciliations and decentralised ledgers.

**m. Organisational Digital Readiness:** Empirical studies [21] confirm that firms with advanced cloud systems and in-house IT security are significantly more prepared to adopt post-quantum tools, including PQC and QKD deployment.

**n. Stakeholder Interdependence:** As finance increasingly involves coordination among accountants, auditors, tax consultants, and regulators, QKD enables secure multi-party transactions and real-time signature verification [7].

### 5.1.2 Proposition 2:

**Integrating quantum computing into accounting practices mitigates risks associated with data breaches and unauthorised access to financial information.**

**a. People:** For example, accountants and IT professionals may need to upskill and reskill to work with quantum computing technologies. Quantum computing has security implications in cybersecurity processes. Integrating quantum computing into practice settings introduces tensions and challenges to accounting and those individuals responsible for cybersecurity. Professionals in accounting and IT must upskill in quantum concepts to safely adopt these technologies. Quantum computing integration causes interdisciplinary tension but opens pathways for secure innovation in financial data handling [57].

**b. Quantum Key Distribution (QKD) Integration:** In addition, integrating QKD into the accounting systems ensures that the encryption keys used to protect financial information are shared via a secure channel that cannot be transmitted to a third party (hider or eavesdropper), which means that an attack has occurred, but the information remains safe. Thus, accounting systems that integrate quantum computing are less vulnerable to attack via malicious programs or human intervention, making financial data more secure. In conclusion, [97] integrates QKD into accounting systems, offering secure communication channels that are accountable, traceable, and security-based in financial transactions. Quantum cryptography and QKD are already being applied in high-security sectors to ensure data integrity and detect eavesdropping, thereby securing accounting systems [98].

**c. Perceived Usefulness:** Implementing quantum algorithms, such as Grover's algorithm, is significantly more secure than all currently available encryption methods and serves as a vital safeguard against infiltration approaches, offering both enhanced security and speed. Quantum Key Distribution (QKD) is precisely what accounting jobs require to secure financial data from ever-evolving cyber threats. It significantly enhances the applicability of quantum computing to accounting and cybersecurity jobs. Studies show that Grover's algorithm and QKD outperform classical encryption and key distribution methods. These are mission-critical safeguards against cyberattacks in data-rich sectors, such as accounting [99].

**d. Complexity:** The software's technical complexity could be a barrier. However, once user-friendly platforms and interfaces are developed, perceived ease of use likely decreases the accountants' learning curve. Researchers acknowledge that technical barriers currently exist. However, trends point to ongoing simplification through standards, platform development, and guided frameworks for integration [57].

**e. Relationship:** Sophisticated new cyberattacks prompt significant industry bodies and financial regulators to endorse the use of quantum technologies to mitigate risks. Quantum encryption techniques, which are based on quantum mechanics, may initially seem intimidating. However, their capacity to enhance security could simplify the administration of cyber threats, making covert attacks more difficult. The global rise in cybercrime, combined with regulations such as GDPR and financial data mandates, drives organisations to seek quantum-secure methods, as seen in parallel sectors like healthcare [98].

**f. Environment:** Cooperation between quantum computing researchers and accounting software vendors can facilitate the integration of quantum solutions. Enhanced data security through quantum encryption can increase stakeholders' confidence in the security of accounting systems. There is documented cooperation between research bodies, accounting scholars, and quantum technologists to create frameworks for quantum-resistant integration into business practices [57].

**g. Attitude Towards Use:** The continuously evolving sophisticated techniques hackers employ to create a positive attitude toward using quantum solutions. Accounting professionals should be open to adopting quantum encryption if they are aware of its benefits. So, the result is a boost in confidence that quantum technology is more secure than classical techniques. The increasing recognition of quantum computing's potential to counter sophisticated cyber threats drives positive perceptions among tech-savvy accounting professionals [100].

**h. Culture:** A culture that prizes innovation and security hastens the diffusion of quantum encryption methods; accountants who see cyber threats are likely to act preemptively. A proactive culture of innovation and compliance in financial firms accelerates the adoption of secure, traceable, and quantum-safe methods for financial record-keeping [101].

**i. Organisational Structure:** If accounting firms have suffered a breach or learned of other firms' deployment of quantum-safe practices, their intention to use quantum-based systems escalates. Integrating quantum computing creates a new interdependency between accountants and IT professionals, who work together to ensure encryption technologies are in place. Historical breach cases and regulatory trends suggest that firms exposed to quantum-safe use cases or targeted by cyberattacks move swiftly toward integration [98].

**j. Perceived Consequences:** Studies such as [5, 7, 102] emphasise the financial and reputational damage resulting from data breaches in accounting systems.

**k. Quantum Accounting Algorithms – Quantum Resistance:** A quantum-resistant algorithm secures the integrity and secrecy of accounting activities, protecting accounting records against potential damage from future quantum threats. Quantum-resistant algorithms safeguard financial data in preparation for the era of quantum computers, making the concept of quantum encryption particularly promising for accounting applications.

### 5.1.3 Proposition 3:

**Adopting quantum computing in accounting requires substantial changes in cybersecurity protocols, but leads to superior financial data protection:**

**a. Technology Implementation:** Adopting quantum computing necessitates a profound overhaul of existing cybersecurity measures. Businesses and governments must complete various tasks, including establishing new quantum-safe encryption standards, upgrading IT infrastructure, and training personnel to utilise quantum technologies. Researchers argue for the industry-wide deployment of quantum-safe cryptography, such as lattice-based systems and post-quantum RSA alternatives [103]. Quantum computing introduces disruptive encryption capabilities requiring firms to redefine cybersecurity architectures—from network protocols to data storage systems. Overhauls include new cryptographic standards, hardware upgrades, and skilled personnel capable of handling post-quantum algorithms. [104].

**b. Task Interdependence:** How financial information is processed and protected also changes. Quantum computing opens up new possibilities for encryption, storage, and data transmission, which must be integrated into accounting workflows to enhance efficiency. Quantum-enhanced systems redefine financial information workflows, especially where encryption, storage, and verification intersect. New protocols require deeper integration with AI-driven anomaly detection and quantum key distribution (QKD)- -based communications [105].

**c. Quantum Resistance of Accounting Algorithms:** This is one of the reasons why we need to develop quantum-proof accounting algorithms. These algorithms would necessitate changes to cybersecurity implementations, resulting in superior protection of a business's financial data on the Internet. Security is enhanced when classical accounting systems are integrated with quantum-resistant encryption, which strengthens protection against future quantum attacks [57].

**d. Perceived usefulness:** Even though we need new protocols to make quantum-safe accounting viable, quantum computing adds enhanced security against cyberattacks and breaches, so accounting wants to use it. So, it would seem sensible for accounting to invest in new, radically secure options, such as quantum computing, to continue fulfilling its roles. Quantum computing enhances the fit for accounting roles by helping the profession move toward enhanced, quantum-safe security, which addresses problems more effectively. Studies affirm that QML, QKD, and lattice-based cryptography offer more secure, traceable, and unbreakable encryption for financial data, making quantum adoption highly attractive for accounting professionals [105].

e. **Complexity:** Implementing it initially feels somewhat challenging, as it involves training and upgrading cybersecurity infrastructure, which is likely to be less than smooth. However, the benefits can be long-term and obvious once in place. Combining quantum technologies can be complicated, but the initial technical work can be completed once the learning curve is overcome. Although initial integration is complex, frameworks and standards are emerging to help firms adopt quantum security protocols with minimal disruption [57].

f. **Organisational Digital Readiness:** Establishing government policies and corporate investment in quantum computing research and development (RD) helps facilitate the transition to quantum-based cybersecurity. IT and accounting professionals must work more closely together on each task, hopefully establishing closer working relationships. Multiple papers advocate for interdisciplinary collaboration between quantum scientists, IT departments, and financial regulators to enable successful cybersecurity transformation [106]. Joint efforts across public-private partnerships support the shift toward quantum cybersecurity [107].

g. **Attitude toward Use:** Over time, with proper training, many accountants have come to embrace these new technologies, which provide an effective layer of protection for their data storage. Perspectives may initially start as either scepticism or fear. Positive attitudes toward quantum cybersecurity are increasing, particularly among firms that experience firsthand the benefits or recognise quantum threats as imminent [108]. Firms that adopt quantum-enhanced protocols report greater confidence in the protection of their financial data [108, 109].

h. **Quantum-resistant algorithms:** Accounting systems must upgrade their interfaces and incorporate quantum-resistant algorithms into these new protocols, thereby making financial data resistant to quantum attacks[110]. Integrating QKD and post-quantum algorithms creates accounting systems that are both mathematically and physically hacker-proof, significantly reducing the probability of breaches [103, 110]. Adoption ensures reduced breach risks and accelerates the uptake of quantum technology [44, 95].

i. **Integration of Quantum Key Distribution (QKD):** QKD provides secure communication channels essential for future accounting cybersecurity. Integrating quantum key distribution is crucial in developing future cybersecurity protocols, as it provides secure communication channels for accounting systems. Ensuring secure communication channels by integrating quantum key distribution (QKD) methods reduces the overall risk of these systems. QKD also require changes in the communication protocol being used, supporting progress in the evolving nature of more secure accounting systems. Quantum key distribution is key to building next-generation accounting communication protocols that are verifiably secure and cannot be eavesdropped upon [103].

j. **Relationship Factors:** Quantum cyber security calls for increased cooperation between accountants and information technology experts [104]. Joint responsibility in encryption facilitates interdisciplinary collaboration [105]. Past cases of breaches call attention to the importance of synergistic cybersecurity governance [98].

### 5.1.4 Proposition 4:

**Challenges such as high costs, the need for specialised skills, and the complexity of quantum computing initially hinder its widespread adoption in the accounting sector. However, the enhanced security benefits eventually outweigh these challenges.**

a. **Job Fit:** Currently, the skills mismatch between accountants and quantum computing results in a job mismatch [111]. Over time, quantum capabilities will be increasingly embedded in accounting tools, enhancing job fit and performance [112].

b. **Environmental Context:** Firstly, the cost and complexity of quantum computing, not to mention the paucity of skilled professionals, are severe impediments to using quantum computing in accounting's usual ways. However, as the technologies concerned mature and become more accessible, is it safe to assume that security gains drive their use despite initial obstacles? Quantum adoption is hindered by high financial costs, lack of talent, and limited infrastructure, especially in developing markets [113]. However, security and computational advantages drive widespread uptake as technologies mature [114].

c. **Organisational Characteristics:** Firms must conduct a cost-benefit analysis when considering whether to invest in quantum computing, should it become available soon [107, 115]. The capital outlay can be significant, but the longer-term benefits of enhanced security – or, more accurately, compliance with forthcoming regulatory standards – are considerably more critical than the initial outlay. Firms are advised to weigh short-term cost hurdles against long-term benefits in terms of compliance, accuracy, and risk mitigation [115]. Adoption is considered a strategic investment, not just a technological one [107].

d. **Quantum Key Distribution (QKD) Integration:** As QKD becomes more reliable and its costs decline, investment in new accounting technology and human resources training ensures that its superior security is widely adopted, thereby overcoming the initial barriers. QKD is initially costly and complex, but it ultimately enables secure, quantum-resilient communication channels, which are critical for sensitive financial systems [116].

e. **Perceived Usefulness:** Though the short-term difficulties are substantial, there ultimately are longer-term benefits in terms of security, which effectively outweigh the inertia headache for quantum computing. In the short term, due to the skills gap, there is an increased likelihood that job fit will worsen, i.e. jobs did not sufficiently provide individuals with the required skills to utilise the technology according to its needs. In the long term, as the role catches up to the apps used in accounting, the job fit typically becomes obsolete again. While current implementation is challenging, long-term cybersecurity and performance gains in finance and accounting are well established [62].

**f. Perceived Ease of Use:** Today, quantum systems are relatively complex to understand, which requires highly skilled workers to operate them, and these skills are also in short supply. These are all barriers, but they decrease over time as quantum tools become available and workers' literacy in quantum matters improves. Complexity presents a notable obstacle to initial adoption, as it is challenging for firms to determine how to integrate quantum technologies into their operations. High initial complexity is one of the most significant barriers to adoption [111]. User-friendly platforms and training programs are needed to lower the entry threshold over time [112].

**g. Environment:** Commercial adopters in the financial and accounting professions, followed by their peers, cascade, reducing the cost of adoption by mitigating the resistance quotient and demonstrating quantifiable value. Resistance to adoption is driven by an aversion to change and fear of complexity [109]. Adoption is slow initially, but peer pressure, industry leadership, and regulatory mandates likely accelerate acceptance and standard-setting [109].

**h. Technology Adoption:** Governments and educational institutions must provide access to quantum computing skills by offering laboratories and study materials and provide further incentives (beyond expected market forces) to drive adoption and help realise the potential benefits. Studies highlight the need for government-led initiatives in quantum education and infrastructure to reduce friction and democratise access [113].

**i. Attitude Towards Use:** Security requirements progressively overcome the entrenched resistance to the costs and complexity of quantum computing. As organisations begin to observe the successes of quantum-enhanced security, resistance to the costs and complexities is reduced, fostering optimism [117].

**j. Organisational Quantum Readiness:** The adoption of quantum computing in accounting increases as firms realise its apparent security benefits and gain a competitive advantage despite resistance to these quantum computing applications due to concerns about costs and complexity. Evident resistance due to costs and complexity may initially be present (or, perhaps more realistically, apathy), but the positive outcomes of early adopters can boost industry confidence and enthusiasm. Companies are already preparing for adoption by investing in research pilots and quantum readiness audits [115].

**k. Professionals, such as accountants and IT specialists,** need to develop a basic understanding of quantum mechanics. Cross-discipline cooperation and upskilling solutions are needed to close this skills gap and enable safe implementation [57]

**l. Culture:** An innovation and cybercrime-aware culture supports the active uptake of quantum technology in pioneering companies. Tech-savvy and compliance-oriented accounting organisations will be most open to uptake [101].

### 5.2 Discussion – Answers to Research Questions:

The findings are summarised from the proposition-based analysis and extensive literature-based theme review, directly answering the study's research questions. The findings reflect how quantum computing and its

Page **33** of **44**

technologies are shaping the present and future cybersecurity status of accounting and finance systems, concerning the theoretical constructs developed in Section 3.

**Research Question 1:** How can quantum computing improve cybersecurity in accounting systems (compared to the classical cryptographic methods)? Quantum computing facilitates quantum-safe cryptographic protocols (lattice-based encryption, QKD) with unbreakable encryption assurance [58, 116]. Shor's algorithm's vulnerability affects classical RSA and ECC algorithms; however, lattice-based cryptography (Kyber, Dilithium) resists and protects accounting records and audit logs [5, 95]. QKD ensures secure key exchange, thereby avoiding man-in-the-middle attacks in accounting data exchanges [118]. Grover's algorithm facilitates faster and more secure searching and verification techniques, enhancing audit procedures [99]. Implementation of these protocols preserves financial integrity, long-term confidentiality, and non-repudiation in sensitive accounting systems [44].

**Research Question 2:** How secure can quantum-secure cybersecurity frameworks prevent unauthorised access to sensitive financial and accounting information? QKD offers provable eavesdrop resistance even in the presence of quantum attackers [98, 118]. Advanced quantum cryptographic systems provide secure multi-party accounting communication layers [7]. Grover's algorithm supports secure authentication and authorisation systems as a means of preventing illegal access [99]. Deploying immutable audit trails via QKD and blockchain systems ensures the integrity of accounting records [32]. Empirical evidence suggests that accounting systems with integrated QKD are more effective in detecting illegal intervention than classical models [103].

**Research Question 3:** What changes must be made in the accounting and finance departments regarding the organizational and structural dimensions to implement quantum cybersecurity systems? Firms need to redefine their cybersecurity governance structures to incorporate quantum-safe protocols [104]. Cross-functional teams must be established, consisting of accounting, IT security, and compliance professionals [112]. Accounting personnel need specific training in quantum cryptography principles and quantum-resistant audit procedures [57]. Business workflows need to be rearchitected to include QKD key management procedures, quantum-safe signature verification, and advanced anomaly detection [105]. A culture of continuous innovation and compliance in cybersecurity will be necessary to build readiness [101]. The implementation of quantum technologies must align with strategic objectives and regulatory mandates [115].

**Research Question 4:** What are the critical adoption barriers to quantum cybersecurity in accounting, and can these be overcome with long-term security gains? Primary obstacles include high upfront costs, a shortage of trained specialists, and technical complexity [113, 114]. The digital readiness of organisations varies greatly; companies with high cloud and IT sophistication overcome these challenges most effectively [21]. Subjective norms and cultural resistance initially hinder adoption, but success stories from pioneering users [115] and social influence [109] ultimately propel widespread industry adoption in the long run. Long-term advantages encompass increased security, compliance with regulations, and a competitive edge [62]. Government initiatives and standardisation activities (e.g., NIST PQC, ISO) contribute significantly to reducing obstacles [106]. The costs of QKD will decrease, and easy-to-use quantum cybersecurity platforms will help bridge the skills gap [116].

## 6.0 Contributions of the study:

### 6.1 Academic Contributions:

The study contributes to the academic understanding of integrating quantum computing in cybersecurity within the context of accounting and finance. It helps bridge the knowledge gap on how quantum computing can be leveraged to enhance data security in accounting and finance at a time when conventional cybersecurity measures are increasingly becoming vulnerable to quantum-based attacks ([117]). In addition, the study enhanced the information system model proposed by Thompson et al. [46] to include quantum computing-based cybersecurity, providing accountancy scholars with new conceptual models to base their research on. This integration of quantum computing technologies in accounting systems aims to maintain confidentiality and system integrity. Adapting the information system model is relevant to academics seeking to identify the opportunities and threats of quantum computing to accounting and finance services, where confidentiality and system integrity are highly required.

### 6.2 Methodological Contributions (use of Psalars):

The study adopted a systematic literature review methodology (PSALSAR), ensuring methodological rigour and repeatability (Mengist et al., 2020b). The PSALSAR methodology, which includes phases such as Research Protocol and Reporting of Results, ensured that the study could synthesise and analyse existing research without bias. This study, therefore, contributes to the methodology of conducting systematic studies. This research demonstrates that PSALSAR can be utilised to investigate emerging technologies, including quantum computing and cybersecurity. The study can provide a benchmark for future studies in emerging interdisciplinary fields, developing a structured approach to synthesising and analysing the literature.

### 6.3 Policy Contributions to Technology Regulations in Accounting:

The study contributed to the development of policy regarding technology regulations in accounting. The study explains how quantum technologies pose regulatory challenges and how existing regulatory frameworks should evolve to accommodate these technologies, thereby enabling their responsible and secure implementation in accounting and finance. Policymakers can utilise the findings from this study to develop guidelines and standards that ensure the regulated and secure adoption of quantum computing in accounting and finance. The study also recommends that policymakers collaborate with financial institutions and researchers to establish regulatory frameworks for quantum technologies that promote innovation and security. The collaborative approach to regulating quantum technologies will facilitate the development of a secure financial infrastructure in the quantum era.

## 7.0 Future Research:

Although several post-quantum cryptographic algorithms are available, few are designed explicitly to handle accounting functions such as audits, ledger reconciliation, or compliance. Future work should focus on designing and testing quantum-resistant algorithms tailored to accounting workflows. Output would include open-source algorithmic frameworks that conform to international cryptographic standards. Integrating quantum technologies into legacy ERP and accounting systems is a challenging task. Work should aim to create interoperability solutions, such as middleware layers that enable hybrid encryption in legacy environments. That affords practical on-ramping roadmaps for real-world implementation. The successful adoption of quantum computing in accounting involves accountants, IT experts, quantum engineers and compliance professionals. Research should propose socio-technical frameworks that apply UTAUT models and organisational change theory to assess quantum readiness. An organisation's quantum readiness toolkit is the expected output. Given the cost concern highlighted in Proposition 4, a financial analysis is necessary to compare the investment in quantum cybersecurity to the risk cost of future breaches. It may involve modelling different scenarios for firms of various sizes. The output may include an executive-level decision-support model for accounting and finance professionals. Few studies simulate real quantum threats, such as Shor's or Grover's algorithm, against financial systems. Work must use quantum simulators to simulate these scenarios and test current defences. Resultant deliverables include a threat simulation framework for financial institutions and auditors. As quantum technologies impact transparency and traceability, their effect on accounting standards warrants further study. Future work should analyse how quantum methods intersect with GAAP, IFRS and SOX compliance. The output includes guidelines for the ethical deployment of quantum tools in regulated environments. A composite index that considers infrastructure, regulatory environment, education and market exposure is needed to benchmark international readiness. The goal is to establish a global Quantum Security Index (QSI) for the finance and accounting sectors.

Future studies should build upon this framework by incorporating empirical validation, such as case studies, interviews, or pilot implementations of quantum-secure accounting systems. Comparative studies across industries could identify sector-specific nuances and best practices. Quantitative assessments of adoption costs, regulatory compliance outcomes, and performance improvements post-QKD or PQC implementation would further enrich this field. Researchers are also encouraged to explore the impact of quantum security on real-time auditing, blockchain integration, and cross-border financial governance.

## 8.0 Conclusion:

In summary, quantum computing enhances the security of accounting practices by providing more robust encryption methods, improving risk management, enabling advanced anomaly detection, and facilitating efficient data processing. Collectively, these capabilities offer a more secure and resilient framework for protecting sensitive financial information than traditional computing methods.

Quantum computing can fundamentally change the parameters applied for AI-based decision-making in OM. It could redefine how to utilise AI based on new parameters regarding the decision search space, size of the alternative set, interpretation of process and outcomes, speed of decision-making, and replicability of processes. More specifically, quantum computing may lead to new decision-making processes in operations management (OM).

An integrated conceptual framework combining user-level, organisational, and quantum-specific constructs was developed and used to evaluate four key propositions. The analysis revealed that quantum computing offers measurable advantages in encryption strength, data integrity, and regulatory compliance but faces adoption challenges related to cost, complexity, and cultural readiness.

Quantum computing represents both a disruptive threat and a transformative opportunity for the accounting profession. As data-driven decision-making becomes increasingly central to financial reporting and control, ensuring the security and resilience of systems becomes paramount. This article emphasises the need for proactive strategic planning, robust training programs, and regulatory foresight to ensure that the benefits of quantum cybersecurity are realised while mitigating associated risks. The integrated framework and research propositions presented here offer a foundation for both scholarly inquiry and practical transformation in the quantum era of accounting.